\documentclass[preprint]{elsarticle}
\usepackage{amsmath,amssymb}
\usepackage{slashed}
\usepackage{graphicx}%
\usepackage{subfigure}
\usepackage{mathptmx}
\usepackage{url}
\newcommand{\beq}{\begin{equation}}
\newcommand{\eeq}{\end{equation}}

\newcommand{\be}{\begin{eqnarray}}
\newcommand{\ee}{\end{eqnarray}}
\long\def\hidestart#1\hideend{}
\begin{document}

\title
{Topological susceptibility in Lattice QCD with unimproved  Wilson fermions}

\author{Abhishek Chowdhury}
\ead{abhishek.chowdhury@saha.ac.in}

\author{Asit K. De}
\ead{asitk.de@saha.ac.in}

\author{Sangita De Sarkar}
\ead{sangita.desarkar@saha.ac.in}

\author{A. Harindranath\corref{cor1}}
\ead{a.harindranath@saha.ac.in}

\author{Santanu Mondal}
\ead{santanu.mondal@saha.ac.in}

\author{Anwesa Sarkar}
\ead{anwesa.sarkar@saha.ac.in}

\address{Theory Division, Saha Institute of Nuclear Physics \\
 1/AF Bidhan Nagar, Kolkata 700064, India}
\author{Jyotirmoy Maiti}
\ead{jyotirmoy.maiti@gmail.com}
\address{Department of Physics, Barasat Government College,\\
10 KNC Road, Barasat, Kolkata 700124, India}

\cortext[cor1]{Corresponding author}
\date{August 6, 2011}
\begin{abstract}
We address a long standing problem regarding topology in lattice simulations
of QCD with unimproved Wilson fermions. Earlier attempt with 
unimproved Wilson fermions 
at $\beta =5.6$  
to verify the suppression of topological susceptibility with decreasing quark 
mass ($m_q$) was unable to unambiguously confirm the suppression. 
We carry out systematic calculations for two 
degenerate flavours at two different lattice spacings 
($\beta = 5.6$ and $5.8$). 
The effects of quark 
mass, lattice volume and the lattice spacing on the spanning of different 
topological sectors are presented. We unambiguously 
demonstrate the  suppression of the topological 
susceptibility with decreasing  quark mass, expected  from chiral Ward 
identity and chiral perturbation theory. 

\end{abstract}

\maketitle
%%%%%%%%%%%%%%%%%%%%%%%%%%%%%%%%%%%%%%%%%%%%%%%%%%%%%%%%%%
\section{Introduction}\label{intro}
%%%%%%%%%%%%%%%%%%%%%%%%%%%%%%%%%%%%%%%%%%%%%%%%%%%%%%%%%%
Because  of the explicit violation of chiral 
symmetry by a dimension five kinetic operator, there have been persistent 
concerns about the Wilson formulation \cite{wilson1, wilson2} of fermions on
the lattice in reproducing the chiral properties 
of continuum QCD. To address these concerns, for the past few years, we have 
been studying \cite{anomaly1, anomaly2} the chiral properties of Wilson 
lattice QCD. We have studied  the emergence of the chiral anomaly with 
unimproved 
and improved Wilson fermions, the associated cutoff effects and  
the approach to the infinite  volume chiral  limit in the context of flavour
singlet Axial Ward Identity to order ${\cal O}(g^2)$. 
In this work we address 
various issues associated with topological 
charge ($Q$) and topological susceptibility ($\chi$) in lattice QCD simulations with two 
degenerate flavours of unimproved Wilson fermions. The detailed account of low lying
spectroscopy and autocorrelation studies will appear separately.

Earlier attempt \cite{bali} with unimproved Wilson fermions 
and HMC algorithm performed simulation at $\beta =5.6$, 
lattice volumes $16^3 \times 32$ and
$24^3 \times 40$
and $m_{\pi}\geq 500$ 
MeV. Their results  as well as the results from other 
collaborations \cite{anna} were presented,  adopting a mass-dependent 
renormalization scheme,
by scaling 
topological susceptibility and $m_{\pi}^2$ by appropriate powers of 
{\em quark mass dependent} 
$r_0/a$ where $r_0$ is the Sommer parameter. Since $r_0/a$  significantly 
increases with decreasing 
quark mass, the 
suppression of topological susceptibility may be 
concealed in such a plot especially for large pion masses and for lattice 
actions which have more severe cutoff artifacts. 
Results of Ref. \cite{bali} 
failed to show the suppression of topological 
susceptibility unambiguously when data was presented in this manner:
 see Fig.12
in Ref. \cite{bali}. (Note however that, at present, a mass-independent 
renormalization scheme 
appears to be preferred in the lattice literature where $r_0$ in the chiral 
limit is used to scale the data.)   
Since topological 
susceptibility is a measure of the spanning of different topological sectors 
of QCD vacuum, the inability to reproduce the predicted suppression may raise 
concerns about the simulation algorithm and the particular fermion formulation 
to span the configuration space correctly. 
Unimproved Wilson fermion has 
$\cal{O}$(a) lattice artifact  and hence it is also important to study 
the effects of scaling violation. 
%So far, studies have been done with 
%pure gauge, improved Wilson fermion, 
%staggered fermion and with chiral fermions. 
%However, no systematic study has 
%  
We perform a systematic study using 
unimproved Wilson fermions at different volumes, different lattice spacings
($\beta$ = 5.6 and 5.8)
and $m_{\pi}\geq 300$ MeV. We present our data in both 
mass-dependent and mass-independent
renormalization schemes.
In the former scheme, our data
(at both couplings $\beta$=5.6 and 5.8) clearly show the suppression of 
topological susceptibility in the region of pion mass $m_\pi \leq $ 500 MeV. 
In the latter scheme, our data (at both couplings $\beta$=5.6 and 5.8) exhibit
 the suppression of 
topological susceptibility in the whole region of pion masses studied. 
When plotted using mass independent scheme, 
the data of Ref. \cite{bali} also clearly
show the suppression of topological susceptibility
for the avaiable range of their data ($m_{\pi} \ge 500$ MeV). 
%and demonstrate the suppression of topological susceptibility with decreasing 
%quark mass. 
Since our data exhibits suppression in the lower pion mass region 
{\em independent of the renormalization scheme used}, 
we are able to  demonstrate {\em unambigously} the suppression of topological 
susceptibility with decreasing quark mass, expected in continuum QCD in the 
region of lower pion mass.
Advancements in both algorithm and technology have made this study possible.

%%%%%%%%%%%%%%%%%%%%%%%%%%%%%%%%%%%%%%%%%%%%%%%%%%%%%%%%%%%%%%%%%%%%
\section{Measurements}
%%%%%%%%%%%%%%%%%%%%%%%%%%%%%%%%%%%%%%%%%%%%%%%%%%%%%%%%%%%%%%%%%%%%
We have generated ensembles of gauge configurations by means of HMC
\cite{hmc,milc} and DDHMC \cite{ddhmc}
algorithm using unimproved Wilson fermion and gauge actions with 
$n_f=2$ mass degenerate quark flavours. At $\beta=5.6$ the lattice volumes are 
$16^3 \times 32$, $24^3 \times 48$ and $32^3 \times 64$ and the renormalized 
physical quark mass (calculated using axial Ward identity)
ranges between $15$ to $100$ MeV ($\overline{\rm MS}$ scheme 
 at $2$ GeV). At $\beta = 5.8$ the lattice 
volume is $32^3 \times 64$ and the renormalized physical quark mass 
 ranges 
from $20$ to $90$ MeV. The lattice spacings determined using Sommer parameter
at $\beta =5.6$ and $5.8$ are $0.077$ and $0.061$ fm respectively. 
The configurations for lattice volumes $24^3 \times 48$ and $32^3 \times 64$ 
and for $16^3 \times 32$ at $\kappa = 0.15775$ are generated using DDHMC
algorithm.
The number of thermalized configurations ranges from $2000$ to $12000$ 
and the number of measured configurations ranges from $70$ to $500$.

For topological charge density, we use the lattice approximation developed for 
$SU(2)$ by DeGrand, Hasenfratz and Kovacs \cite{degrand}, modified for
$SU(3)$ by Hasenfratz and Nieter \cite{hasenfratz1} and implemented in
the MILC code \cite{milc}. 
It uses ten link paths
described by unit lattice vector displacements in the sequence $\{x,y,z,-y,-x,
t,x,-t,-x,-z\}$ and $\{x,y,z,-x,t,-z,x,-t,-x,-y \}$ plus rotations and cyclic
permutations.  
To suppress the ultraviolet lattice artifacts, smearing of link fields 
is required. The link field is smeared by $20$ HYP smearing steps
and optimized smearing coefficients $\alpha =0.75$,
$\alpha_2=0.6$ and $\alpha_3=0.3$ \cite{hasenfratz2}.   
We have observed that smearing brings the topological charges close to 
integer values
and behaviour is better for the smaller lattice spacing, as expected.
\begin{figure}
\includegraphics[width=4in,clip]{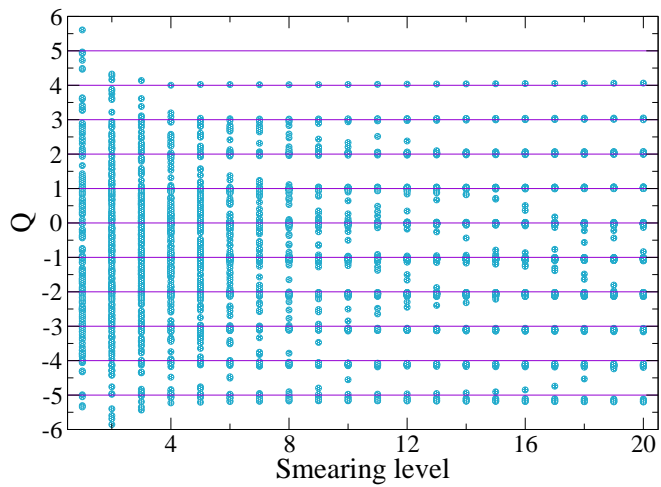}
\caption{Topological charge of gauge configurations 
versus HYP smearing steps for $\beta = 5.8$ and 
$\kappa = 0.15475$ ($m_{\pi} \sim 300 {\rm MeV}$) at lattice volume $32^3 \times 64$. }
\label{fig1}
\end{figure}

\begin{figure}
\includegraphics[width=4in,clip]{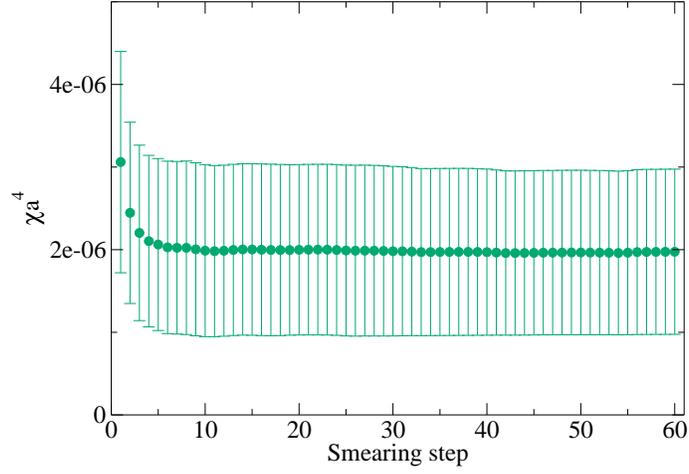}
\caption{Topological susceptibility  
versus HYP smearing steps for $\beta = 5.8$ and 
$\kappa = 0.15475$ at lattice volume $32^3 \times 64$. }
\label{fig2}
\end{figure}

In Fig. \ref{fig1} we show the behaviour of topological charge of gauge 
configurations with smearing steps for $\beta = 5.8$ and 
$\kappa = 0.15475$ at lattice volume $32^3 \times 64$. It is evident that 
the topological charges 
of different configurations are clustering about the  integer values 
after about $10$ smearing steps and values are stable under further 
smearing steps.
In Fig. \ref{fig2} we present the behaviour of topological susceptibility
with smearing steps for the same lattice parameters which shows that the 
susceptibility is very stable with smearing steps after $10$ steps.

\begin{figure}
\subfigure{
\includegraphics[width=2.5in]{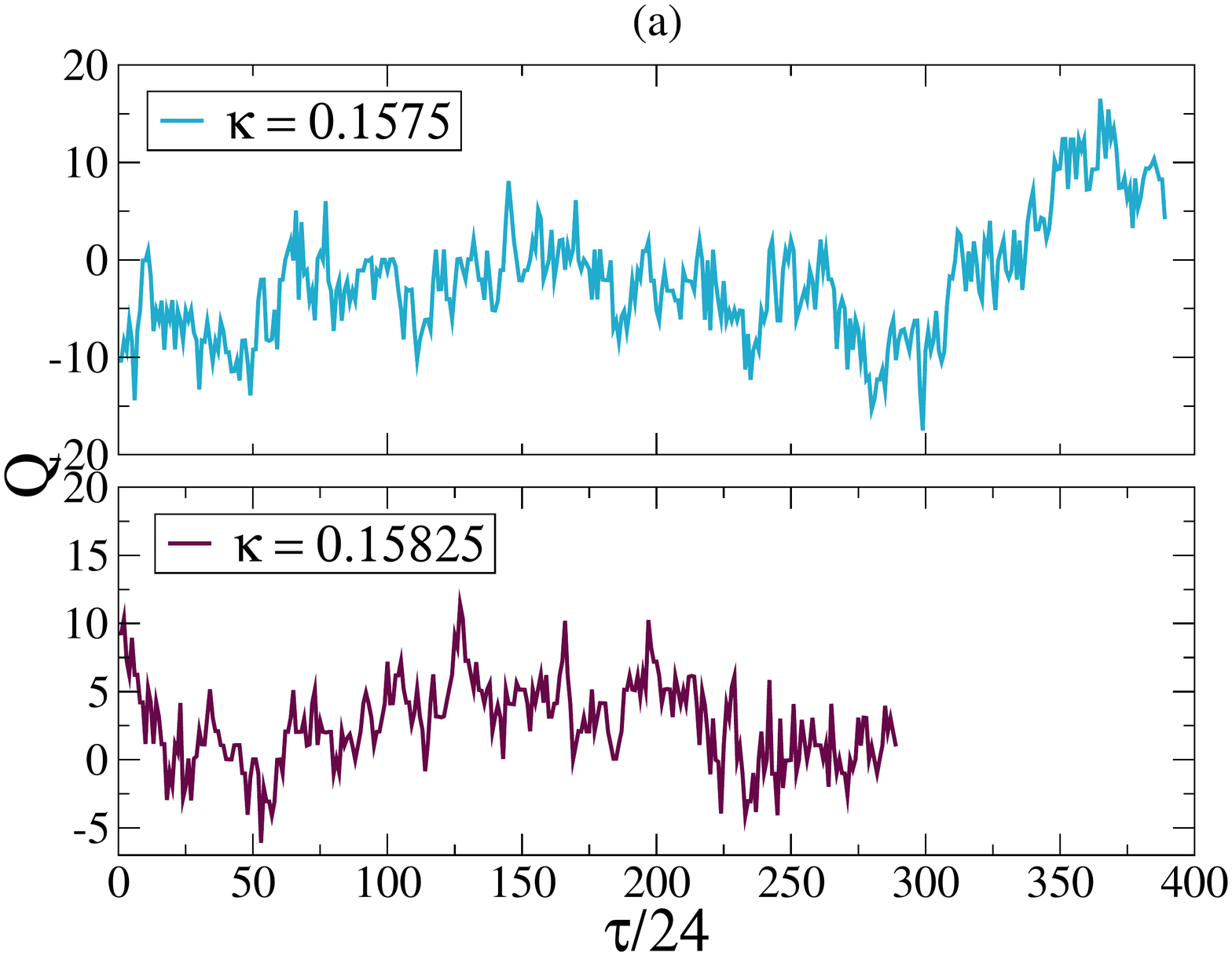}
}
\subfigure{
\includegraphics[width=2.5in]{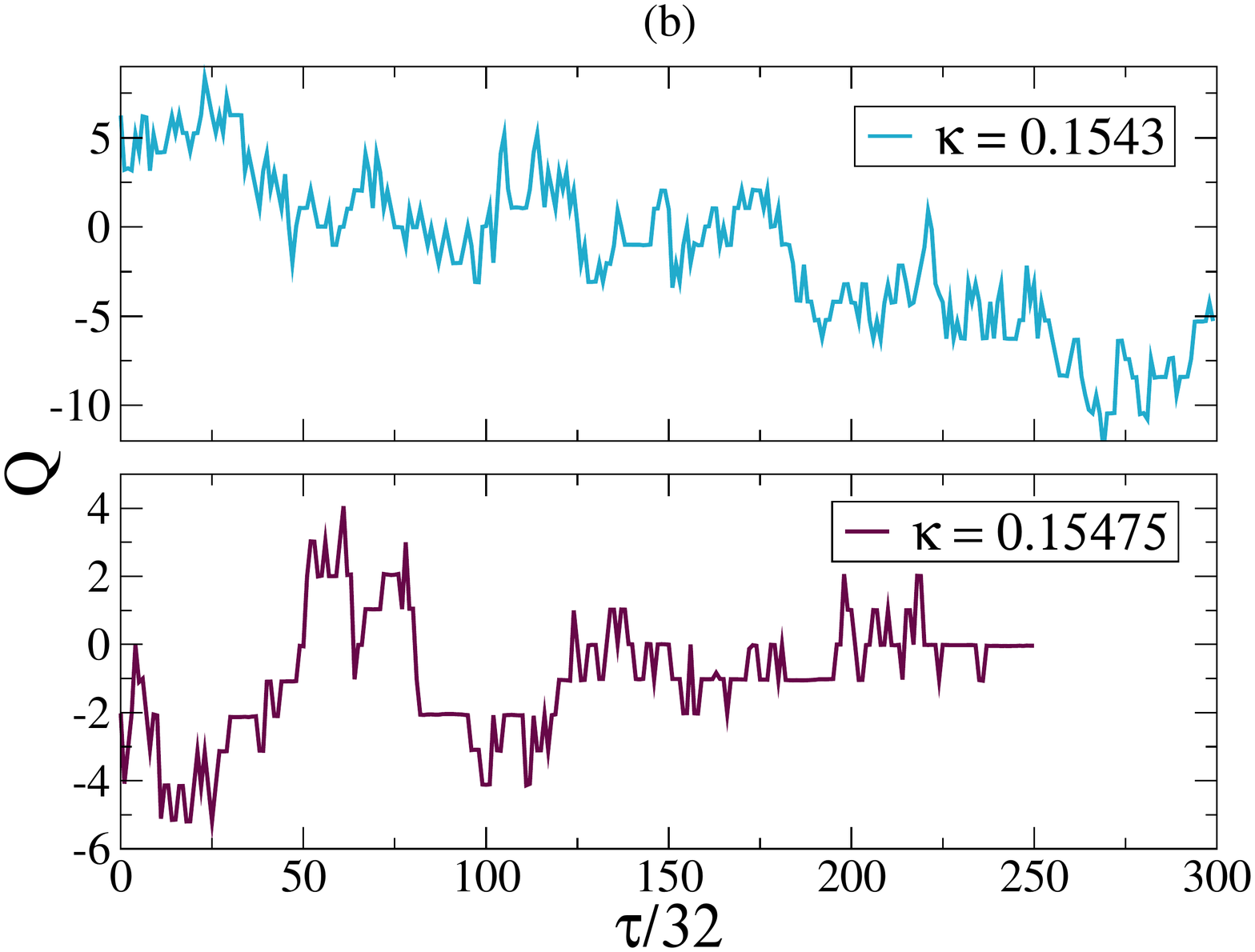}
}
\caption{The Monte Carlo trajectory history for topological charge with 
unimproved Wilson fermion and gauge action for (a) $\beta = 5.6$ 
with a gap of 24 trajectories between two consecutive measurements  
and 
(b) $\beta = 5.8$ 
with a gap of 32 trajectories between two consecutive measurements.}
\label{fig3}
\end{figure}

In Fig. \ref{fig3} we show the Monte Carlo time history of topological charge 
for $\beta = 5.6$ and $5.8$ for the smallest and the largest $\kappa$ and 
there is 
some evidence of trapping of the topological charge only at $\beta = 5.8$ and 
largest $\kappa$. 

\begin{figure}
\subfigure{
\includegraphics[width=2.5in]{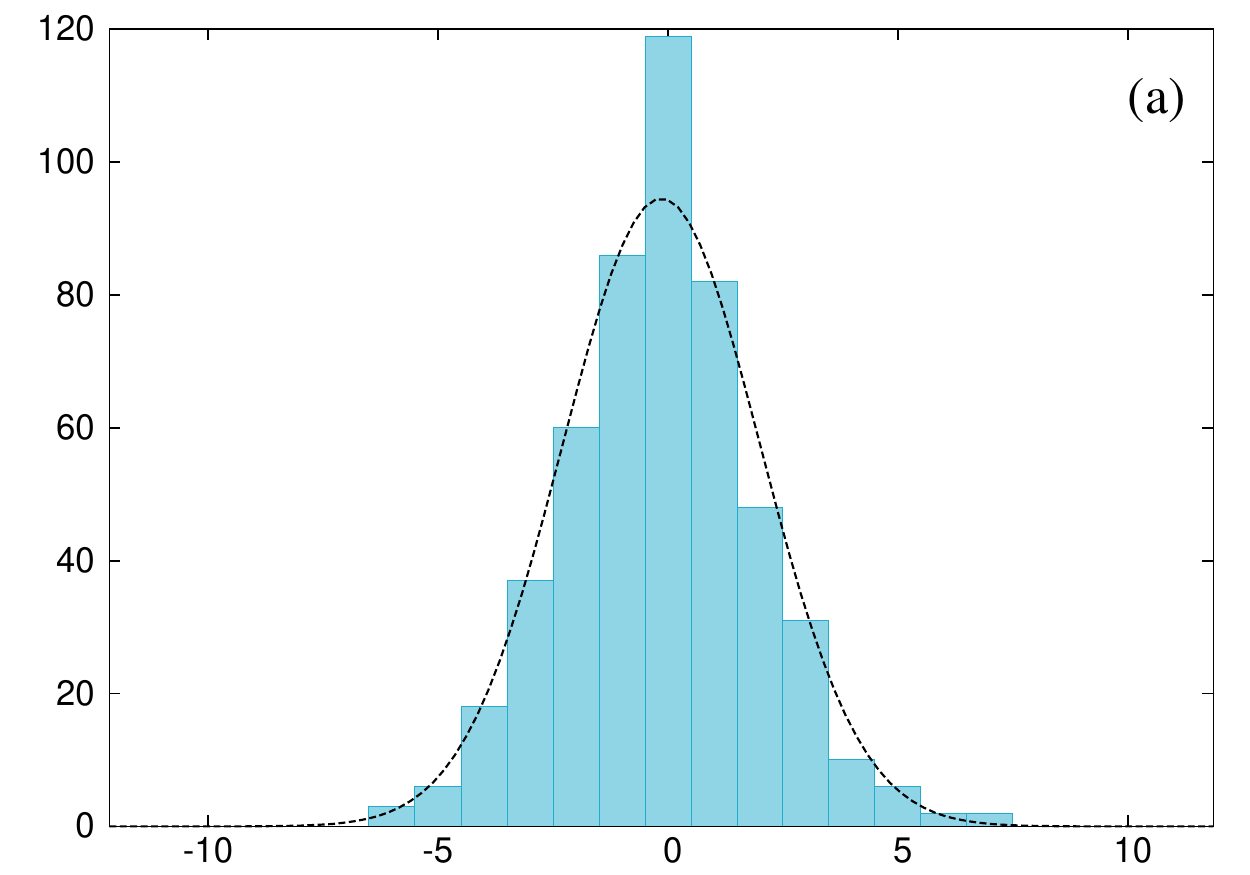}
%\label{fig3}
}
\subfigure{
\includegraphics[width=2.5in]{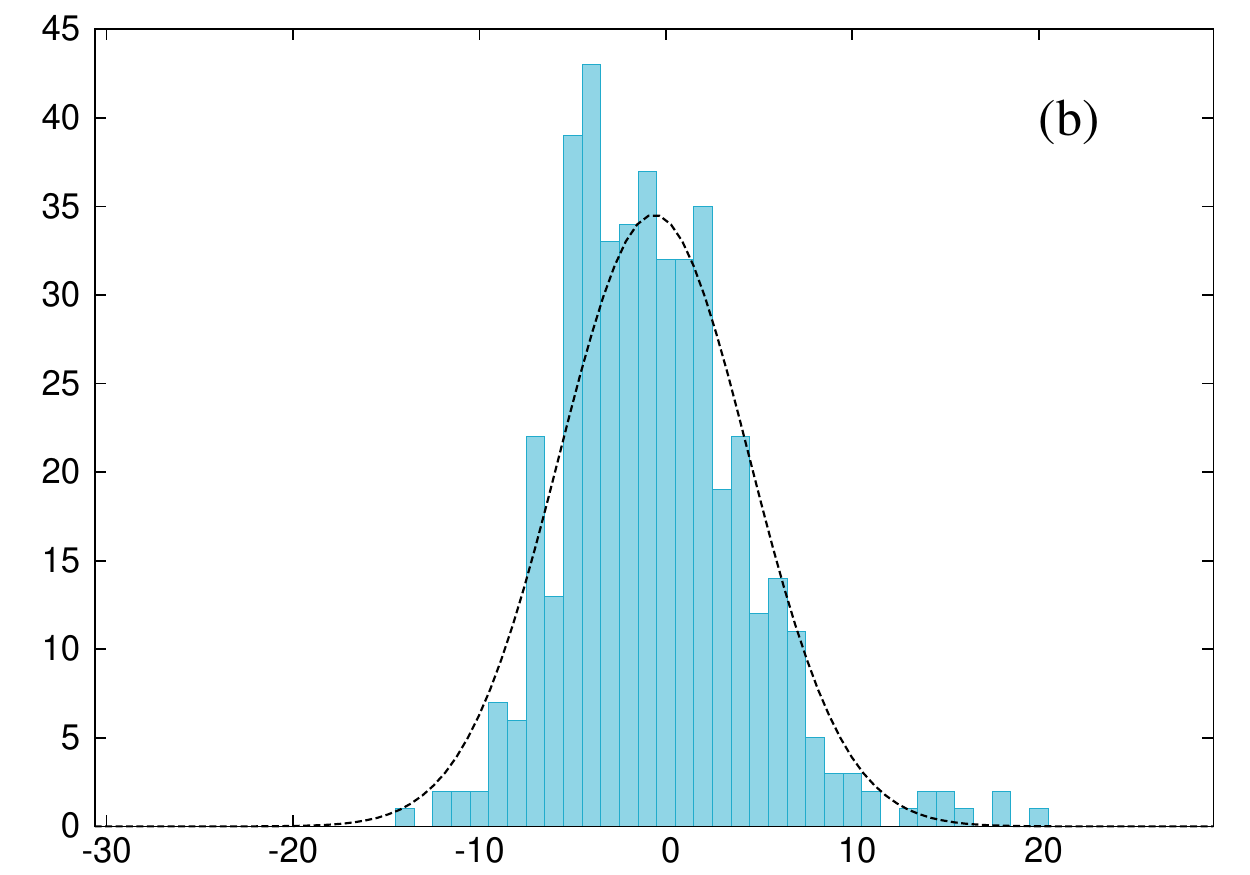}
}

\subfigure{
\includegraphics[width=2.5in]{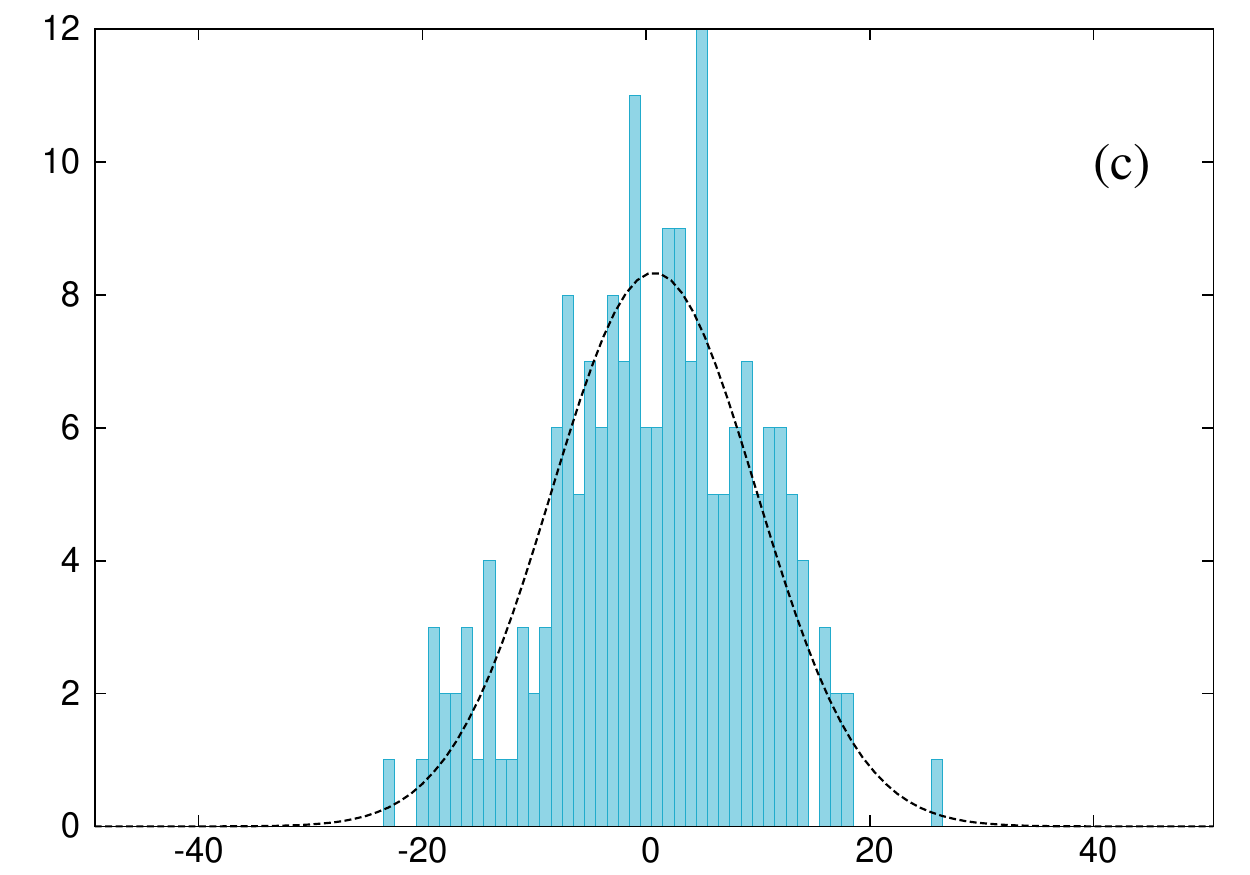}
}
\subfigure{
\includegraphics[width=2.5in]{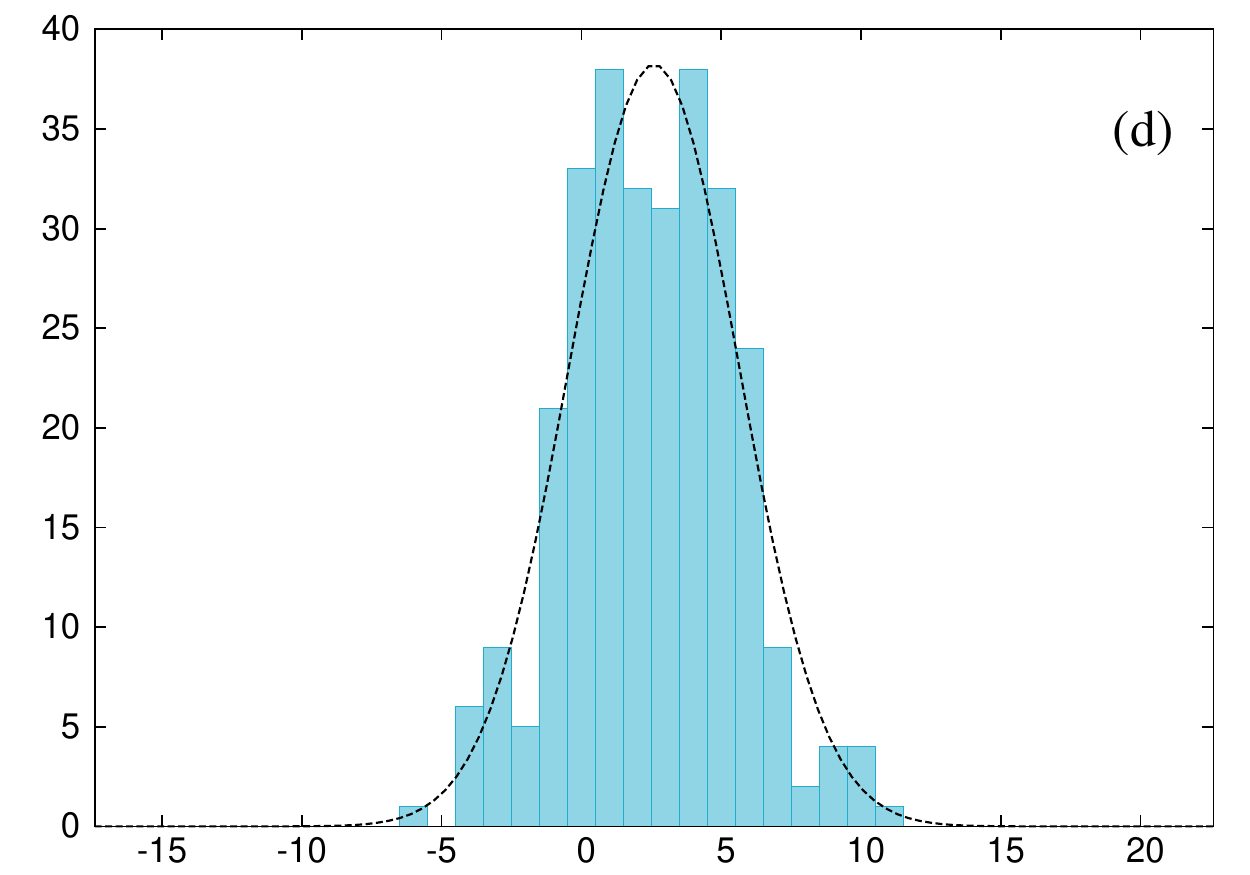}
}
\subfigure{
\includegraphics[width=2.5in]{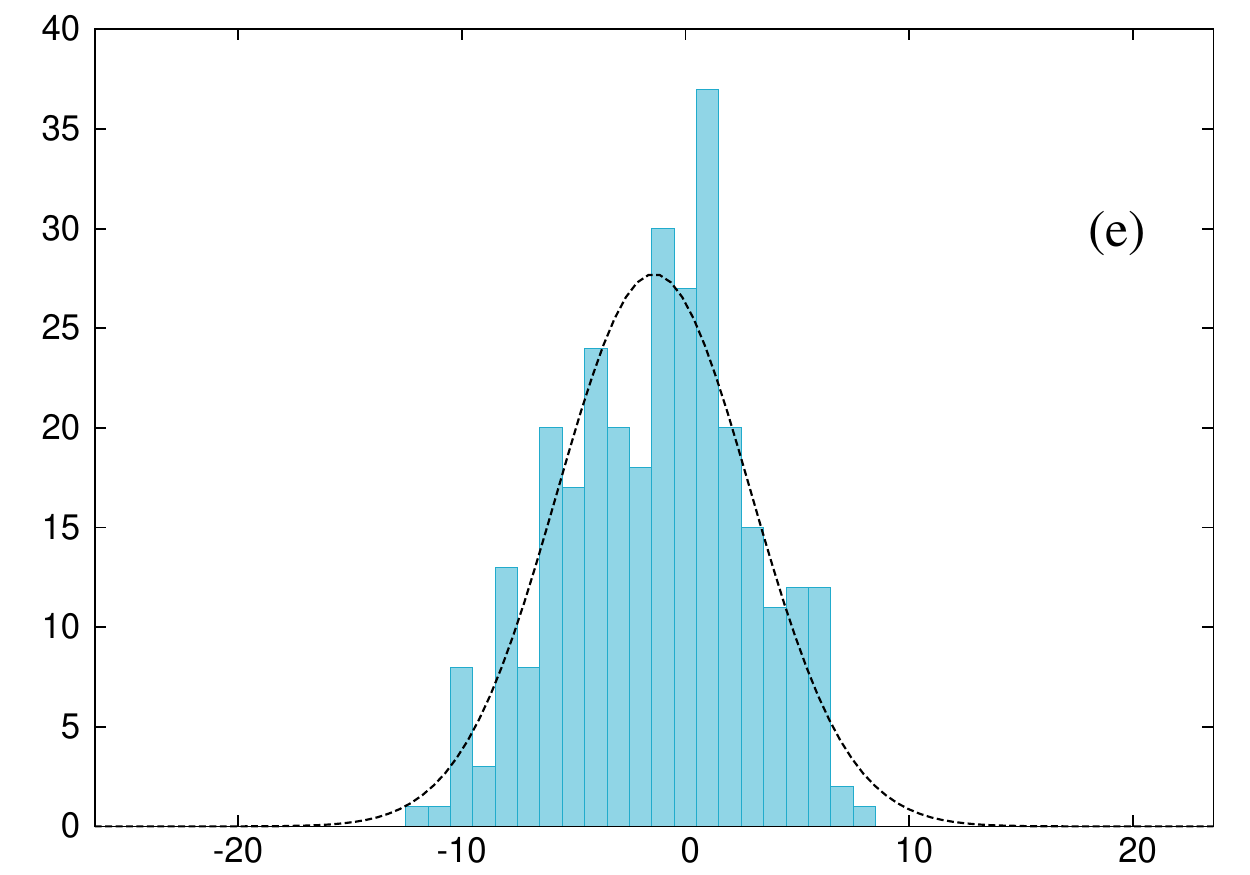}
}
\subfigure{
\includegraphics[width=2.5in]{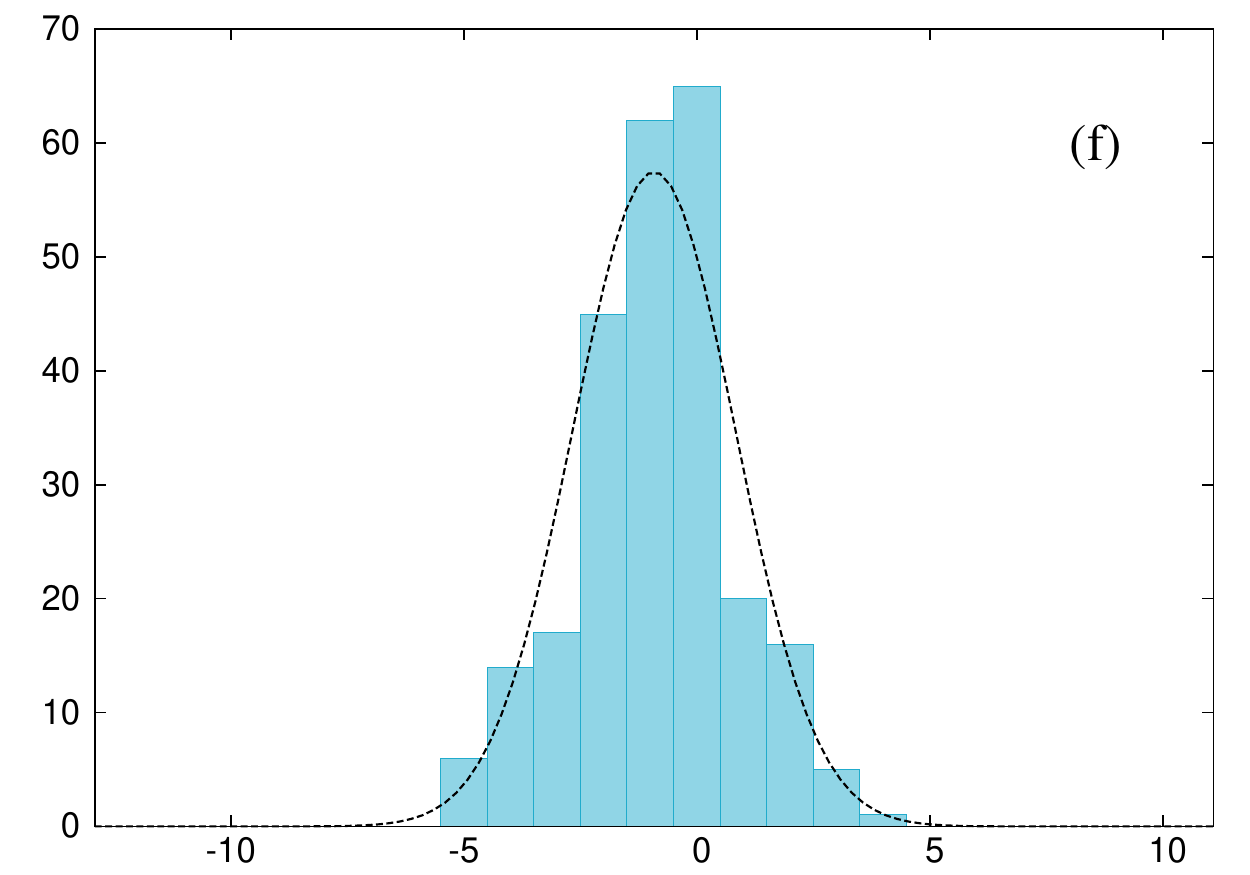}
}
\caption{The topological charge distribution for 
(a) $\beta = 5.6$, $\kappa = 0.15775$, volume $=16^3\times 32$
(b) $\beta = 5.6$, $\kappa = 0.15775$, volume $=24^3\times 48$
(c) $\beta = 5.6$, $\kappa = 0.15775$, volume $=32^3\times 64$
(d) $\beta = 5.6$, $\kappa = 0.15825$, volume $=24^3\times 48$
(e) $\beta = 5.8$, $\kappa = 0.1543$, volume $=32^3\times 64$
(f) $\beta = 5.8$, $\kappa = 0.15475$, volume $=32^3\times 64$.
}
\label{fig4}
\end{figure}

Fig. \ref{fig4} displays six histograms of topological charge distributions,
for two values of $\beta$ and different volumes. 
The topological charge data were put in several bins and 
the bin widths were chosen to be 
unity centered around the integer values of the topological charges for all 
the cases. From theoretical considerations the distribution of the topological 
charge is expected to be a Gaussian \cite{gaussian}. Since 
our configurations are large in number but finite, an incomplete 
spanning of the topological sectors may occur and    
$\langle Q \rangle$ may  not be zero. Hence we define the susceptibility 
to be 
\be
\chi = \frac{1}{V}\left(\langle Q^2\rangle-\langle Q \rangle^2\right).
\label{eq1}
\ee
The expected distribution is
\be
n_Q=\frac{n_{meas}}{\sqrt{ 2\pi(\langle Q^2\rangle-\langle Q \rangle^2)}}
exp\left(-\frac{Q^2}
{2(\langle Q^2\rangle-\langle Q \rangle^2)}\right)\label{eq2}
\ee
where $n_{meas}$ is the total number of measurements made. 
The Gaussian curves in Fig. \ref{fig4} are obtained by using Eq. \ref{eq2}. 
It is evident from Fig. 
\ref{fig4} that for a given $\beta$ and volume the width of the distribution 
decreases as $\kappa$ increases, indicating that the topological susceptibility
$\chi$ decreases with decreasing quark mass.

%%%%%%%%%%%%%%%%%%%%%%%%%%%%%%%%%%%%%%%%%%%%%%%%%%%%%%%%%%%%%%%%%%
\section{Results}
%%%%%%%%%%%%%%%%%%%%%%%%%%%%%%%%%%%%%%%%%%%%%%%%%%%%%%%%%%%%%%%%%
\begin{table}
\begin{center}
\begin{tabular}{|l|l|l|l|l|l|}
\hline 
\hline
$\beta=5.6$ & \multicolumn{5}{c|}{} \\
%&\multicolumn{2}{c|}{$am_\rho$}&
%\multicolumn{1}{c|}{$aF^{V}_\rho$}\\
%\cline{2-4}
\hline
 &$lattice$& $\kappa$&  $am_{\pi}$ & {$am_q$} &{$\chi a^4/10^{-5}$} \\
\hline
&{$16^3\times 32$}&{$0.156$} &{0.4456(19)} &{0.06724(78)}&{11.9020(1.255)}\\
&{$,,$}&{$0.157$} &{0.3441(27)}&{0.04004(61)}&{7.6147(0.948)}\\
&{$,,$}&{$0.1575$} &{0.2867(27)}&{0.02813(57)}&{5.6381(0.863)} \\
&{$,,$}&{$0.15775$} &{0.2529(28)}&{0.02018(50)}&{3.6700(0.919)} \\
&{$,,$}&{$0.158$} &{0.2258(35)}&{0.01388(48)}&{2.9405(0.354)} \\
&{$24^3\times 48$}&{$0.1575$} &{0.2717(15)} &{0.02669(37)}&{5.4727(1.56)} \\
&{$,,$}&{$0.15775$} &{0.2387(21)} &{0.02127(37)}&{3.8986(0.764)}\\
&{$,,$}&{$0.158$} &{0.1967(24)} &{0.01441(29)}&{3.9070(0.800)}\\
&{$,,$}&{$0.158125$} &{0.1767(23)} &{0.01143(22)}&{2.6140(1.30)}\\
&{$,,$}&{$0.15825$} &{0.1478(24)} &{0.00690(17)}&{1.3804(0.646) }\\
&{$32^3\times 64$}&{$0.15775$} &{0.2359(16)} &{0.02042(29))}&{3.9826(0.574}\\
&{$,,$}&{$0.158$} &{0.1979(16)} &{0.01489(32)}&{3.5056(0.775)} \\
&{$,,$}&{$0.15815$} &{0.1669(17)} &{0.01070(19)}&{2.4784(0.829)}\\
&{$,,$}&{$0.1583$} &{0.1301(20)} &{0.00632(18)}&{1.4420(0.678)}\\
\hline \hline
\hline
$\beta=5.8$ & \multicolumn{5}{c|}{} \\
\hline
 &$lattice$& $\kappa$&  $am_{\pi}$ & {$am_q$} &{$\chi a^4/10^{-5}$} \\
\hline
&{$32^3\times64$}&{$0.1543$} &{0.1585(11)} &{0.01338(14)}&{0.8891(0.248)}\\
&{$,,$}&{$0.15455$} &{0.1228(10)} &{0.00767(9)}&{0.3704(0.144)}\\
&{$,,$}&{$0.15475$} &{0.0920(23)} &{0.00365(11)}&{0.1447(0.087)} \\
\hline \hline
\end{tabular}
\end{center}
\caption{Pion masses, unrenormalized quark masses and topological susceptibilities
in lattice units.}
\label{lattice}
\end{table}
\begin{figure}
\includegraphics[width=4in,clip]{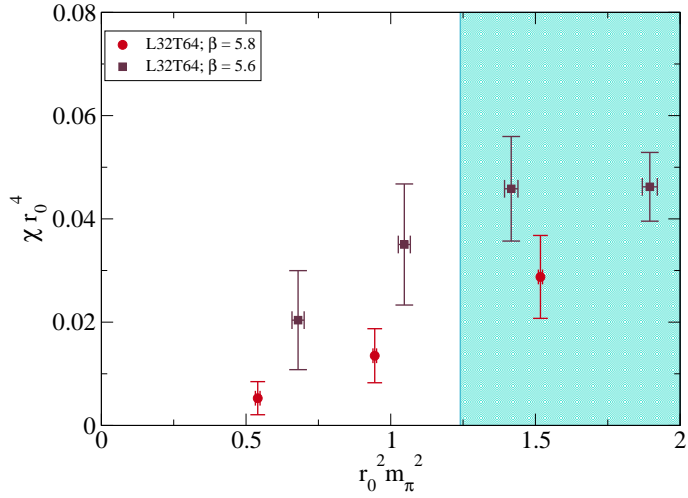}
\caption{Topological susceptibility  
versus $m_{\pi}^2$ in the units of $r_0$ (quark mass dependent) 
for $\beta = 5.6$ and $5.8$
and at lattice volume $32^3 \times 64$. }
\label{fig5}
\end{figure}

\begin{figure}
\includegraphics[width=4in,clip]{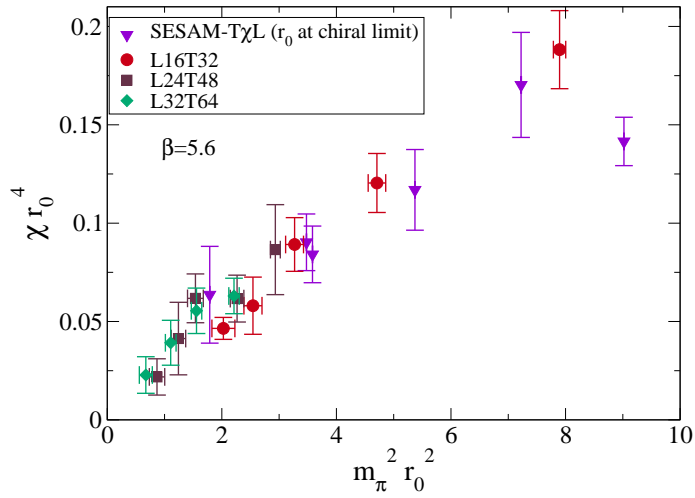}
\caption{Topological susceptibility  
versus $m_{\pi}^2$ in the units of $r_0$ (at chiral limit) for $\beta = 5.6$ 
and at lattice volumes $16^3 \times 32$, $24^3 \times 48$, and $32^3 \times 64$
compared with the results of SESAM-T$\chi$L collaborations \cite{bali}. }
\label{fig6}
\end{figure}
\begin{figure}
\centering
{\includegraphics[width=4in,clip]{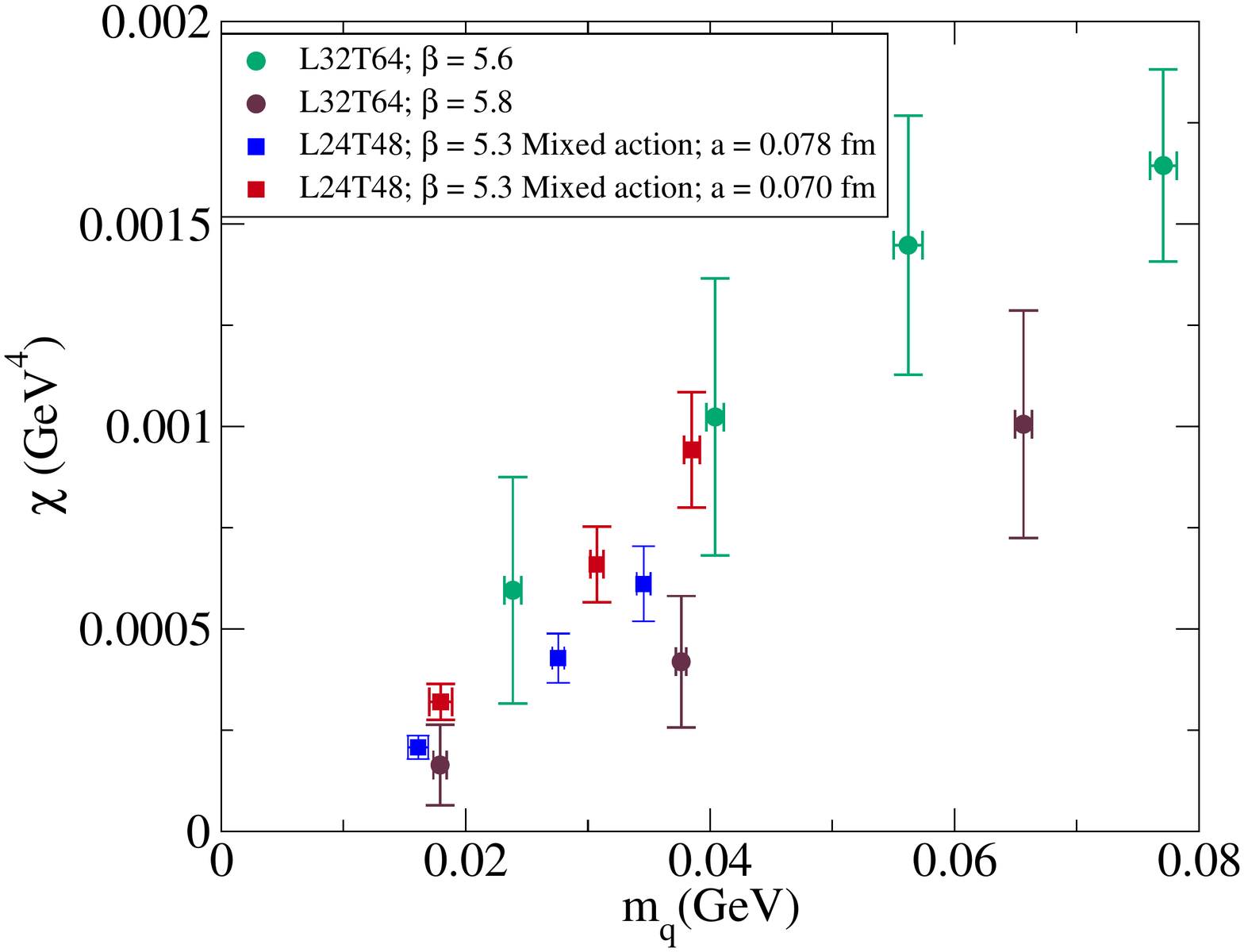}
\caption{Topological susceptibility  
versus $m_{q}$ in the physical units  for $\beta = 5.6$ and $\beta = 5.8$
 at lattice volume $32^3 \times 64$
compared with the results of mixed action (Clover and Overlap) 
\cite{bernardoni}.  For the latter, the two separate lattice spacing
determinations,
 $a =  0.0784$ fm and 
$a =0.070$ fm quoted have been used. }
\label{fig7}}
\end{figure}
In table \ref{lattice} we present, in lattice units,  the pion masses, 
unrenormalized quark masses and topological susceptibilities measured.
For ease of comparison with earlier presentations of results and 
 clarification
 we present  for $\beta = 5.6$ and $5.8$
 at lattice volume $32^3\times 64$, topological susceptibility  
versus $m_{\pi}^2$ in the units of $r_0$ (quark mass dependent), 
 (i.e., in mass-dependent renormalization scheme)
 in Fig. \ref{fig5}.
 At $\beta =5.6$ our second largest quark mass (in Fig. \ref{fig5}) is 
very close to the
lowest quark mass of Ref. \cite{bali}. In this figure the shaded region
corresponds to $m_{\pi} \geq 500$ MeV. In the lower pion mass region 
our data for both $\beta = 5.6$ and $5.8$ clearly show the suppression
even in mass-dependent renormalization scheme. 
  
 Fig. \ref{fig6} shows our results for topological susceptibility versus 
$m_{\pi}^2$, in the units of Sommer parameter ($r_0$) at the 
chiral limit (i.e., in the mass-independent renormalization scheme), 
for $\beta = 5.6$ and
 at lattice volumes $16^3 \times 32$, $24^3 \times 48$, and $32^3 \times 64$.
We also show the results of SESAM-T$\chi$L collaborations \cite{bali}.
%Our error bars incorporate the integrated autocorrelation times for all 
%quantities.
 Using 
the numbers given in \cite{bali},  we have replotted it after scaling by the 
value of $r_0$ quoted at the physical point. This new plot clearly shows the 
suppression of susceptibility with decreasing quark mass in  the earlier
SESAM-T$\chi$L data with unimproved Wilson fermion. Our results carried out
at larger volume and smaller quark masses unambiguously 
(i.e., independent of the renormalization schemes used)
establish the 
suppression of topological susceptibility with decreasing 
quark mass in accordance with
the chiral Ward identity and chiral perturbation theory. 
Further we note that, for a given $\kappa$, the value of topological 
susceptibility increases with the volume as expected from finite 
volume considerations. 
This effect is more noticeable 
in smaller volumes.
For large enough volume topological susceptibility should be independent 
of volume, since $\langle Q^2 \rangle$ scales with the volume.       

Since unimproved Wilson fermion has $\cal{O}$(a) lattice artifacts
it is important to estimate the scaling violations of our results.
Fig. \ref{fig7} shows  topological susceptibility
in the physical units  for $\beta = 5.6$ and $\beta = 5.8$
 at lattice volume $32^3 \times 64$  
versus nonperturbatively renormalized \cite{becirevic}
 quark mass in $\overline{\rm MS}$ scheme \cite{gimenez} 
at $2$ GeV. Topological susceptibility at $\beta =5.8$ approximately matches 
with that
at $\beta= 5.6$ within the error bars but the latter data is
 systematically below
the former. This behaviour which is qualitatively consistent with leading
order lattice artifact \cite{veneziano} is 
 observed also by the MILC 
collaboration \cite{bazavov}.  
  For comparison, in Fig. \ref{fig7}, 
we have also shown the results of mixed action calculation of Ref.
\cite{bernardoni}
at $\beta =5.3$ where the authors 
 have quoted  two separate lattice spacings,
 $a =  0.0784$ fm and 
$a =0.070$ fm for the same $\beta$. 
The calculation of Ref. \cite{bernardoni} employs gauge configurations 
generated with dynamical $\cal{O}$(a) improved Wilson fermion, using
DDHMC algorithm and topological charge is measured using a fermionic operator,
namely, the Neuberger-Dirac 
operator. 
Fig. \ref{fig7} shows that
our results of the topological susceptibility favourably
compare with that of Ref. \cite{bernardoni}.
%The lattice spacing has been
%determined to be $a =  0.0784(10)$ fm in \cite{} but, preliminary
%result from more precise determination
%through different methods yielded $a =0.070$ fm \cite{}.  
In Ref. \cite{latt2011} we have shown that (see Fig. 6 of Ref. \cite{latt2011})
 the topological susceptibility at $\beta =5.8$ is consistent with 
leading order chiral perturbation theory prediction.  

In conclusion, we have addressed a long standing problem regarding 
topology in lattice simulations
of QCD with unimproved Wilson fermions. Calculations are presented for two 
degenerate flavours. 
The effects of quark 
mass, lattice volume and the lattice spacing on the spanning of different 
topological sectors are presented. 
By performing simulations at smaller pion masses we have shown 
the suppression of the topological 
susceptibility with respect to  decreasing quark mass irrespective of the 
method of presenting the data. The suppression  
expected  from chiral Ward 
identity and chiral perturbation theory is thus demonstrated unambiguously.
Furthermore, our results of the topological susceptibility favourably
compare with that of $\cal{O}$(a) improved Wilson fermion. 
Our results unequivocally
show that lattice QCD with
naive Wilson fermions together with DDHMC algorithm, for the range of
quark masses and lattice spacings studied, is able to span the configuration
space correctly and  
reproduce the chiral behaviour of continuum QCD. In future work, we would
like to explore lighter quark masses and smaller lattice spacings.
To achieve this goal,
however, improvements in the algorithm might be necessary.
  
\vskip .05in
{\bf Acknowledgements} 
\vskip .05in
  Numerical calculations are carried out on Cray XD1 and Cray XT5 systems 
supported 
by the 10th and 11th Five Year Plan Projects of the Theory Division, SINP under
the DAE, Govt. of India. We thank Richard Chang for the prompt maintenance of 
the systems and the help in data management. This work was in part based on 
the public lattice gauge theory codes of the 
MILC collaboration \cite{milc} and  Martin L\"{u}scher \cite{ddhmc}.

%%%%%%%%%%%%%%%%%%%%%%%%%%%%%%%%%%%%%%%%%%%%%%%%%%%%%%%%%%

%%%%%%%%%%%%%%%%%%%%%%%%%%%%%%%%%%%%%%%%%%%%%%%%%%%%%%%%%%%%%%%%%%%%%%%%   

%%%%%%%%%%%%%%%%%%%%%%%%%%%%%%%%%%%%%%%%%%%%%%%%%%%

\end{document}